\documentclass[twocolumn]{aastex6}

\usepackage{epsfig,graphicx,amssymb,epstopdf,mathrsfs,float,url}
\newcommand{\nufom}{\mbox{$N_{\nu,{\rm obs}}/N_{\nu,{\rm exp}}$}}
\newcommand{\enu}{\varepsilon_\nu}
\newcommand{\egam}{\varepsilon_\gamma}

\newcommand{\veritas}{VERITAS}
\newcommand{\ice}{IceCube}

\def\arcdeg{\hbox{$^\circ$}}

\def\simlt{\mathrel{\hbox{\rlap{\hbox{\lower4pt\hbox{$\sim$}}}\hbox{$<$}}}}
\def\simgt{\mathrel{\hbox{\rlap{\hbox{\lower4pt\hbox{$\sim$}}}\hbox{$>$}}}}

\newcommand{\xray}{\mbox{X-ray}}
\newcommand{\gray}{\mbox{$\gamma$-ray}}
\newcommand{\grays}{\mbox{$\gamma$-rays}}

\newcommand{\Ltev}{\mbox{$L_{\rm TeV}$}}
\newcommand{\Ftev}{\mbox{$F_{\rm TeV}$}}
\newcommand{\phtpercm}{photons\,$\cm^{-2}$}

\newcommand{\percmsqs}{cm$^{-2}$\,s$^{-1}$}

\newcommand{\perval}[2]{{#1\mbox{$^{#2}$}}}

\newcommand{\persec}{\perval{\rm s}{-1}\/}
\newcommand{\percm}{\mbox{$\cm^{-2}$}}

\newcommand{\erg}{\mbox{$\rm\,erg$}\/}
\newcommand{\cm}{\mbox{$\rm\,cm$}}

\newcommand{\cgslum}{\erg\,\persec}

\begin{document}

\title{Search for Blazar Flux-Correlated TeV Neutrinos in
  IceCube 40-String Data}

\author{C.~F. Turley\altaffilmark{1,3,5},
        D.~B. Fox\altaffilmark{2,3,4},
        K. Murase\altaffilmark{1,2,3,4},
        A. Falcone\altaffilmark{2,3},
        M. Barnaba\altaffilmark{2},
        S. Coutu\altaffilmark{1,3},
        D.~F. Cowen\altaffilmark{1,2,3}, \\
        G. Filippatos\altaffilmark{1,3},
        C. Hanna\altaffilmark{1,2,3},
        A. Keivani\altaffilmark{1,3},
        C. Messick\altaffilmark{1,3},
        P. M{\'e}sz{\'a}ros\altaffilmark{1,2,3,4},
        M. Mostaf{\'a}\altaffilmark{1,2,3}, \\
        F. Oikonomou\altaffilmark{1,3},
        I. Shoemaker\altaffilmark{1,3},
        M. Toomey\altaffilmark{1,3},
        G. Te{\v{s}}i{\'c}\altaffilmark{1,3} \\
        (The AMON Core Team)
}

\affil{$^1$Department of Physics, Pennsylvania State University,
           University Park, PA 16802, USA \\
       $^2$Department of Astronomy \& Astrophysics, Pennsylvania
           State University, University Park, PA 16802, USA \\
       $^3$Center for Particle \& Gravitational Astrophysics,
           Institute for Gravitation and the Cosmos, Pennsylvania
           State University, University Park, PA 16802, USA \\
       $^4$Center for Theoretical \& Observational Cosmology,
           Institute for Gravitation and the Cosmos, Pennsylvania
           State University, University Park, PA 16802, USA}

\altaffiltext{5}{\href{mailto:cft114@psu.edu}{\tt cft114@psu.edu}}

\slugcomment{ApJ submitted}
\shorttitle{Search for Blazar Flux-Correlated Neutrinos}
\shortauthors{Turley et al.}


\begin{abstract}
We present a targeted search for blazar flux-correlated high-energy
($\enu\simgt 1$\,TeV) neutrinos from six bright northern blazars,
using the public database of northern-hemisphere neutrinos detected
during ``IC40'' 40-string operations of the IceCube neutrino
observatory (April 2008 to May 2009).
Our six targeted blazars are subjects of long-term monitoring
campaigns by the VERITAS TeV \gray\ observatory. We use the
publicly-available VERITAS lightcurves to identify periods of excess
and flaring emission. These predefined intervals serve as our ``active
temporal windows'' in a search for an excess of neutrinos, relative to
Poisson fluctuations of the near-isotropic atmospheric neutrino
background which dominates at these energies.
After defining the parameters of an optimized search, we confirm the
expected Poisson behavior with Monte Carlo simulations prior to
testing for excess neutrinos in the actual data. We make two searches:
One for excess neutrinos associated with the bright flares of
Mrk~421 that occurred during the IC40 run, and one for excess
neutrinos associated with the brightest emission periods of five other
blazars (Mrk~501, 1ES~0805+524, 1ES~1218+304, 3C66A, and W~Comae), all
significantly fainter than the Mrk~421 flares.
We find no significant excess of neutrinos from the preselected blazar
directions during the selected temporal windows. We derive
90\%-confidence upper limits on the number of expected flux-associated
neutrinos from each search. \added{These limits are consistent with
  previous point-source searches and Fermi GeV flux-correlated
  searches.} \replaced{The}{Our} upper limits are sufficiently close
to the physically-interesting regime that we anticipate future
analyses using already-collected data will either constrain models or
yield discovery of the first blazar-associated high-energy neutrinos.
\end{abstract}


\maketitle


\section{Introduction}
\label{sec:intro}

The \ice\ collaboration has instrumented a cubic kilometer of
Antarctic ice at the South Pole with 86 ``strings'' of photomultiplier
tubes, readout electronics, and an internal calibration system so as
to detect high-energy ($\enu\simgt 1$\,TeV) neutrinos from atmospheric
cosmic ray showers and extraterrestrial sources
\citep{icfacility}. Operations of the 79-string and full-strength
86-string facilities since 2010 recently yielded discovery of a
population of cosmic neutrinos
\citep{ic-pev,Ic3+13-sci,Ic3+14-tevpev}, validating the experiment's
original motivation and approximate scale \citep{wb97,bw01}.  Given
the $\simgt$10\arcdeg\ cascade-type (and less frequently,
$\sim$1\arcdeg\ track-type) positional uncertainties and $\approx$1
per month rate of detection for the highest-confidence cosmic events,
the absence of any demonstrated association with known \gray\
bursts \citep{Ic3+13-sci,Ic3+14-tevpev,icgrb14}, and the inferred
near-isotropic sky distribution \citep{Ic3+13-sci}, the nature of the
sources of these cosmic neutrinos remains uncertain and subject to
active debate (e.g., \citealt{lbd+13,abc+14,murase14,waxman15}).

Ultimately, identification of electromagnetic (EM) counterparts to
high-energy neutrino sources can proceed by one of two possible paths:
Sufficiently variable sources may exhibit correlated variability in
their neutrino and EM emissions; or sufficiently bright individual
sources may (with time) be identified as \textit{a priori} interesting,
astrophysically-plausible EM sources that are coincident with a
point-like or extended excess of neutrinos on the sky.

With respect to the first approach, \gray\ bursts (GRBs) have long been
considered promising candidates for high-energy neutrino emission
\citep{wb97,gkm13,bbm+15}.  Given the brief timescale for high-energy
emission from the typical GRB, and the precision timing of \ice-detected
neutrinos, known GRBs can be readily associated with individual neutrino
events even with the limited positional resolution available for the
\ice\ cosmic neutrinos (mostly cascade events) or associated lower-energy
neutrinos that interact as track events. The relatively relaxed
requirements for positional determination also allow use of the full set
of detected GRBs from Swift, Fermi, the Interplanetary Network, and other
satellites. In spite of this, no coincident high-energy (cascade or track)
or low-energy track neutrino interactions have been found in association
with any detected GRB \citep{icgrb12,antares_grb13}. The inferred limit on
the fraction of the cosmic neutrinos that are due to GRBs is $<$1\%
\citep{icgrb14}.

Without the brief temporal window provided by GRBs, identification and
confirmation of the neutrino-emitting source population(s) must depend on
positional coincidence (particularly via likely-cosmic track-type events)
and careful evaluation of the expected flux and spectrum under proposed
theoretical models. Proposed source populations currently include
star-forming galaxies \citep{lwstarburst2006,mal13,tamstarform2014,smm+15,waxman15},
galaxy clusters and groups \citep{kin08,mal13},
active galactic nuclei (AGN;
\citealt{steckeragn1991,hz97,ad01,muraseagn2014,tg15}), and
quenched-jet GRBs \citep{mw01,mi13,meszarosgrb2014}.
In addition, it remains possible that a minority of the cosmic neutrinos
originate in our own Milky Way Galaxy, either via
the Fermi bubbles \citep{cabubble2011,2014PhRvD..90b3010A,cbcfermibubble2014,lrtbubble2014},
Galactic compact binary TeV \gray\ sources
\citep{milagro12_cygnus,AGPestimate2014} or TeV unidentified sources
(such as pulsar wind nebulae or possible hypernova remnants;
\citealt{fkm13,Budnik+08hn}). All of these scenarios are reviewed by
\citet{murase14}.

The \gray\ bright AGN known as blazars represent a particularly
intriguing possible source population, as they are the most
\gray\ luminous AGN known and the brightest extragalactic sources on
the sky at TeV energies, capable of outshining all other TeV sources
during some flares. Their \gray\ emission is commonly explained by
invoking a leptonic scenario, in which highly-relativistic electrons
in the blazar jets generate emission by a combination of synchrotron
self-Compton and external inverse Compton processes. This scenario
yields minimal high-energy neutrino emission.

However, if or when most of the blazars' high-energy \grays\ are
produced by hadronic interactions in the relativistic jets (with a
minority contribution from leptonic emission processes), the
lepto-hadronic scenario, some or most of the observed \grays\ will
result from cascade emission induced by \grays\ from $\pi^0$
decays. In this case, the observed TeV \grays\ will be accompanied
by comparable fluxes of $\enu\simgt 1$\,TeV neutrinos from decays of
coproduced $\pi^\pm$
\citep{mannheim93,ghsrev95,blazar-model2,czb+15}. In a variant of this
lepto-hadronic scenario, \grays\ are produced by proton synchrotron
radiation without associated high-energy neutrinos \citep{ahar00}.

In another alternative possibility, the intergalactic scenario, the
\grays\ result from intergalactic cascades triggered by
electron-positron pairs that are produced by ultrahigh-energy cosmic
rays escaping from the sources, and photohadronic interactions with
the extragalactic background light then lead to a comparable flux in
high-energy \grays\ and high-energy neutrinos
\citep{ekk+10,mdt+12}.

Significant neutrino emission is not anticipated in the leptonic
scenario \citep{mgcleptonic92,blazar-model2}, nor in the proton
synchrotron version of the lepto-hadronic scenario. However, in a
general sense we expect protons as well as electrons to be accelerated
in blazar jets, and if these AGN are sources of ultrahigh-energy cosmic
rays, associated high-energy neutrino emission can be detectable even
if the observed \grays\ are dominated by leptonic processes
\citep{muraseagn2014}.  Carefully-designed searches for neutrino
emission from the brightest blazars thus have the potential to
constrain source models and important related hypotheses via either
detections (possibly providing the first confirmed counterparts for
any high-energy astrophysical neutrinos) or upper limits.

While leptonic models have been argued to provide a more natural fit
to typical blazar spectra \citep{blazar-model2}, there are unresolved
theoretical challenges. For example, blazar \gray\ flares can show
fast variability, implying a compact emission region which is hard to
reconcile with the non-observation of any high-energy cutoff from
$\gamma\gamma\rightarrow e^+e^-$ \citep{tbg+11,dmt12}. The leptonic
scenario also struggles to explain the occasional ``orphan'' TeV
blazar flares -- flaring emission that is observed at TeV energies
without accompanying flaring in the \xray\
\citep{blottcherorphan05,orphan-flare}.  Even if prototypical blazar
emissions are leptonic, it is still possible that their flares (or
perhaps, their orphan TeV flares) reveal a transient and variable
contribution from hadronic processes.
The June~2002 orphan flare of the blazar 1ES~1959+650
\citep{orphanmulti,sosorphan2013} is of particular note in this
context. \ice's predecessor \mbox{AMANDA-II} detected a
single coincident neutrino from the direction of 1ES~1959+650 during
the flare, although the \textit{a posteriori} nature of the
observation has made it impossible to retrospectively assign a firm
confidence level to any claim of association or non-association
\citep{hh05,sosorphan2013}.
Also worth noting is the recent 9.5~month-long GeV outburst of
PKS~B1424$-$418, which allowed this blazar to dominate the GeV
$\gamma$-ray emissions of all catalogued blazars within the
50\%-containment region of the $\enu\approx 2$~PeV ``HESE 35''
\ice\ event when that neutrino was detected, prompting the suggestion
that this neutrino was emitted by PKS~B1424$-$418 (estimated p-value
of 5\%; \citealt{pev2016}).


Between April~2008 and May~2009, \ice\ ran in a partially-completed
``IC40'' 40-string configuration.  The point-source analysis of this
dataset found no evidence for any neutrino point sources bright enough
to be distinguished above the atmospheric background
\citep{IC-40}. Analysis of data gathered in \ice's subsequent ``IC59''
59-string configuration (May~2009 to May~2010) and further searches
with the full array similarly revealed no point sources, placed upper
limits on the neutrino emissions of an array of candidate sources, and
reported a null result for both a northern-sky time-dependent search
and a Fermi-LAT flux correlated search
\citep{icpoints2015,icfouryears}.


The present work extends previous point-source analyses of the IC40
data by explicitly considering the TeV \gray\ behavior of
the bright northern blazars which are prime candidates for a first
neutrino source detection by \ice. Modeling the TeV lightcurves of
these blazars allows us to select a set of active temporal windows and
blazar directions which are (under our assumed conditions) optimal for
detection of a neutrino excess over anticipated atmospheric
backgrounds. These optimized windows amount in all
cases to less than 11\% of the full IC40 data collection period,
allowing for the possibility of discovering associated neutrino
emission below the threshold of previous time-integrated searches.

Our paper proceeds as follows: We review our datasets and analysis
approach in Sec.~\ref{sec:data_analysis}, and present and discuss our
results in Sec.~\ref{sec:discuss}. We conclude and address possible
future efforts in Sec.~\ref{sec:conclude}.


\section{Datasets and Analysis}
\label{sec:data_analysis}

\subsection{Datasets}
\label{sub:data}

The IceCube 40-string dataset (hereafter IC40) was collected between
April 2008 and May 2009, a total live time of 375.5 days \citep{IC-40}
during which the detector had 40 of the final 86 planned strings
deployed. The data (publicly released in September 2011) records a
total of 12,877 upgoing neutrino events. Extensive analyses
\citep{IC-40,icpoints2015} have constrained the intensities of any
existing point sources. The \ice\ collaboration investigated the
accuracy with which they could reconstruct neutrino arrival directions
using the cosmic ray shadow of the moon \citep{icmoonshadow2013},
and found their reconstruction uncertainty was $0.7^{\circ}$ for the
40- and 59-string configurations.

The six blazars studied in this work are listed with their basic
properties in Table~\ref{tab:blzinfo}. Publicly-available blazar data
(presented in Table~\ref{tab:lcurve}) were collected by the
four-telescope Very Energetic Radiation Imaging Telescope Array System
(VERITAS; \citealt{veritas2008}) and (for Mrk~421 historical data
only) its predecessor, the Whipple Telescope \citep{whip2007}. Both
facilities consist of atmospheric Cherenkov telescopes located at the
Fred Lawrence Whipple Observatory in Arizona, yielding similar
effective energy ranges of $0.1\,{\rm TeV}\simlt \egam \simlt 30\,{\rm
  TeV}$ \citep{wbbveritas2002}, with the exact limits depending on
individual observation and analysis parameters such as elevation angle
in the sky, analysis parameter cuts, and source spectra.  All relevant
blazar data are publicly available from the
\veritas\ website\footnote{Mrk~421 long-term data \citep{14ymrk421}
  are available at
  \url{http://veritas.sao.arizona.edu/veritas-science/mrk-421-long-term-lightcurve}.
  Data for the other blazars are from
  \url{http://veritas.sao.arizona.edu/veritas-science/veritas-blazar-spectra}
      \added{and were not public at the start of this work. The
        data were initially provided for collaborative use and were
        published prior to the completion of this work.}}
Publicly-available historical TeV \gray\ data for Mrk~501 were also
used\footnote{Mrk~501 public light curve data \added{presented in
    \citet{ltcurves} and} available at
  \url{http://astro.desy.de/gamma_astronomy/magic/projects/light_curve_archive/index_eng.html}.}.


\floattable
\begin{deluxetable*}{lrrrrrrrrr}
\tabletypesize{\scriptsize}
\tablecolumns{10}
\tablecaption{VERITAS Northern Blazars \label{tab:blzinfo}}
\tablehead{
 \colhead{Name} &
 \colhead{R.A.} &
 \colhead{Dec.} &
 \colhead{$z$} &
 \colhead{\Ltev} &
 \colhead{\Ftev} & 
 \colhead{$\varepsilon_{\rm th}$}&
 \colhead{$\rm \Gamma_{obs}$}&
 \colhead{$\rm \Gamma_{src}$}&
 \colhead{$N_{\rm obs}$} \\
 \colhead{} & \colhead{} & \colhead{} & \colhead{} & 
 \colhead{($10^{44}$\,\cgslum)} & \colhead{($10^{-11}$\,\percmsqs)} & \colhead{(TeV)}&
 \colhead{} & \colhead{} & \colhead{}
}
\startdata
1ES 0806+524  & 08:09:59.0 & +52:19:00 & 0.138    & 0.1 & 0.2 & 0.3 & 3.6 & 2.7 &  4 \\
1ES 1218+304  & 12:21:26.3 & +30:11:29 & 0.182    & 1.6 & 1.8 & 0.2 & 3.1 & 2.2 & 20 \\
3C 66A        & 02:22:41.6 & +43.02:36 & 0.41\phn & 5.8 & 1.3 & 0.2 & 4.1 & 2.0 & 17 \\
Mrk 421       & 11:04:19.0 & +38:11:41 & 0.031    & 0.4 & 4.6 & 0.4 & 2.2 & 2.0 & 93 \\
Mrk 501       & 16:53:52.2 & +39:45:37 & 0.034    & 0.1 & 3.1 & 0.3 & 2.7 & 2.5 & 15 \\
W Comae       & 12:21:31.9 & +28:13:59 & 0.102    & 0.4 & 1.9 & 0.2 & 3.8 & 3.0 & 19 \\
\enddata

\tablecomments{Coordinates are provided in J2000; $N_{\rm obs}$ gives
  the number of observations during the IC40 observing run. Typical
  luminosities \Ltev\ and fluxes \Ftev\ are over $0.2\,{\rm TeV} <
  \egam < 30\,{\rm TeV}$. Power-law photon indices for Mrk~421 and
  Mrk~501 are known to be variable. The redshift of 3C~66A is
  uncertain, but bounded between 0.33 and 0.44. $\varepsilon_{\rm th}$
  is the threshold energy for VERITAS observations. $\Gamma_{\rm obs}$
  is the measured $\gamma$-ray spectral index, while $\Gamma_{\rm
    src}$ is the spectral index corrected for extragalactic background
  light absorption using the models of \citet{ebl1} and
  \citet{eblinoue}. Blazar properties via TeVCat \citep{tevcat}.}

\end{deluxetable*}


\begin{deluxetable}{ccrrrr}
\tabletypesize{\scriptsize}
\tablecolumns{6}
\tablecaption{Blazar Observations \label{tab:lcurve}}
\tablehead{
 \colhead{~}&
 \colhead{~}&
 \colhead{~} &
 \multicolumn{3}{c}{\Ftev} \\ \cline{4-6}
 \colhead{Name}&
 \colhead{MJD}&
 \colhead{Exp.} &
 \colhead{Obs.} & 
 \colhead{Unc.} &
 \colhead{Rev.} \\
 \colhead{~}&
 \colhead{(d)}&
 \colhead{(h)} &
 \multicolumn{3}{c}{($10^{-11}$\,\percmsqs)}
}
\startdata
1ES 0806+524  & 54564.179 & 0.226 & 2.1\phn & 1.5\phn &  \nodata \\
 ''           & 54566.177 & 2.554 & $-$0.47 & 0.72 & 0.34 \\
 ''           & 54567.153 & 0.562 & 1.0\phn & 1.6\phn &  \nodata \\
 ''           & 54821.408 & 0.595 & 0.52 & 0.83    &  \nodata \\
1ES 1218+304  & 54829.516 & 1.032 & 0.92 & 0.56 &  \nodata \\
 ''           & 54830.524 & 0.869 & 3.11 & 0.67 &  \nodata \\
 ''           & 54838.514 & 1.037 & 1.90 & 0.56 &  \nodata \\
 ''           & 54839.518 & 0.859 & 2.14 & 0.63 &  \nodata \\
 ''           & 54856.459 & 0.298 & 3.3\phn & 1.1\phn &  \nodata \\
 ''           & 54862.454 & 3.072 & 3.79 & 0.37 &  \nodata \\
\enddata 
\tablecomments{Fluxes over $0.2\,{\rm TeV} < \egam < 30\,{\rm TeV}$
  (\Ftev) are reported as observed (Obs.)  with Gaussian uncertainties
  (Unc.), and with forced-positive revised flux estimates (Rev.) where
  necessary. Derivation of revised fluxes is described in
  Sec.~\ref{sub:lightcurves}. Mrk~421 data were obtained from
  \url{http://veritas.sao.arizona.edu/veritas-science/mrk-421-long-term-lightcurve}
  \citep{14ymrk421}. Data for the other blazars are from
  \url{http://veritas.sao.arizona.edu/veritas-science/veritas-blazar-spectra}.
  Table~\ref{tab:lcurve} is published in its entirety in
  machine-readable format. A portion is shown here for guidance
  regarding its form and content. }
\end{deluxetable}

\subsection{TeV Lightcurves}
\label{sub:lightcurves}

We wish to carry out a targeted search for excess neutrinos when the
monitored blazars are relatively bright in TeV \grays. This requires
defining temporal windows of interest for each blazar.  While
\veritas\ makes discrete measurements, the IC40 data were taken
continuously, meaning that every detected neutrino must be judged
either in or out of sample. We choose to make this judgment by
reference to interpolated light curves, which allows us to expand the
temporal windows of interest for each blazar beyond the intervals of
active \veritas\ observation.

This requires us to choose an interpolation approach. We find
traditional linear or spline-interpolation approaches not useful, as
they result in unrealistic predictions during large data gaps and also
(relatedly) because they lack relevant physical underpinning. Since
our primary concern is to avoid accepting neutrinos when the blazar
flux is low, we seek a conservative interpolation that will revert to
a low ``baseline'' flux in the absence of constraining data, thereby
excluding these periods from our detection windows.

\added{Morever, given the infrequent and non-continuous nature of the
  \gray\ measurements, we do not find Bayesian Blocks-type approaches,
  as adopted for example by \citet{flareclass}, to be appropriate in
  this case.} \replaced{At the same time, if we can}{Rather, we seek
  to} extend observed periods of flaring and excess emission in a
realistic and physically-motivated way forward and backward in time,
we will be able to accept neutrinos as potential signal when they are
detected close to these periods. These concerns lead us to adopt a
Gaussian process regression (GPR) \citep{gpml} as our interpolation
approach.

Gaussian Process Regression assumes that process values are generated
as a linear combination of inputs with a multivariate normal
distribution described by stationary mean and covariance functions. It
is a non-parametric method that fits data without choosing a specific
functional form; see \citet{gpml} for details. As applied by us to
\veritas\ blazar observations, the GPR works in two passes. In the
first pass, it analyzes the data to determine the best-fit mean and
covariance ``hyperparameters.'' In our case, we precondition the
blazar light curves, as described below, so that the mean function is
zero, constant and fixed, for each blazar; hence we fit for the
covariance hyperparameters only. On the second pass, the GPR uses the
hyperparameters and observed data values (ignoring the uncertanties)
to predict the maximum likelihood flux (and confidence intervals on
that flux) at all intermediate points. Beyond a few correlation
lengths, the maximum likelihood flux reverts to the mean value, and
the 90\%-confidence bounds expand to encompass the full range of the
observed data, illustrating the fundamentally conservative nature of
this approach.

Our light curve preconditioning begins by replacing negative flux
measurements, which are unphysical, with forced-positive replacement
values (Table~\ref{tab:lcurve}). These values are calculated from the
associated \veritas\ flux estimate and uncertainty as follows:
Treating the flux estimate and uncertainty as the mean and standard
deviation of a Gaussian distribution, we exclude the negative-flux
portion of the distribution and use the median of the resulting
positive-flux distribution as our replacement value.

Next, we calculate a ``baseline flux'' $F_{\rm base}$ for each
blazar. Under the assumption that measurement uncertainties dominate
systematic effects at low flux levels, and that the blazar emission
consists of a dominant baseline flux supplemented by occasional
periods of excess and flaring emission, we define the baseline flux
for each blazar as the mode of the Gaussian kernel density estimator
(KDE) for the blazar flux measurements. The KDE mode should be robust
to a positive ``tail'' of excess and flaring emission episodes. We use
the \texttt{scipy} \citep{scipy} statistics package implementation. The
resulting baseline fluxes are listed in Table~\ref{tab:gpresult}.

In the final step of preconditioning, we divide the blazar flux
measurements (and replacement values) by the baseline flux for that
blazar and take the natural logarithm. Working in log-space reduces
the heteroskedasticity of the flux measurements (which exhibit
substantial positive excursions, but have negative excursions bounded
to the physical range of non-negative fluxes), making the light curves
better suited for GPR analysis. This approach allows us to set the GPR
mean to a fixed constant zero for all blazars.

To carry out the GPR hyperparameter fitting and interpolation, we use
the python package {\tt pyGPs} \citep{pygps}. We run the code using
the squared exponential kernal with isotropic distances, Gaussian
likelihood, exact inference, and a constant mean function, set to
zero. The code optimizes the fit by minimizing the negative log
marginal likelihood with respect to the hyperparameters. The
hyperparameters for Mrk~421 and Mrk~501 were optimized using the
extensive historical data. These hyperparameters were then used to
generate the light curves from the data concurrent with IC40. All
resulting best-fit hyperparameter values are presented in Table
\ref{tab:gpresult}.


\begin{deluxetable*}{lrrrrrrr}
\tabletypesize{\scriptsize} 
\tablewidth{3.2in}
\tablecolumns{8}
\tablecaption{Results of Gaussian Process Regression\label{tab:gpresult}}
\tablehead{
 \colhead{} & \colhead{} & \colhead{} & \colhead{} & 
 \multicolumn{2}{c}{Correlation}&
 \colhead{Noise} & \colhead{} \\ \cline{5-6}
 \colhead{Blazar Name}&
 \colhead{$N_{\rm obs}$}&
 \colhead{$F_{\rm max}$}&
 \colhead{$F_{\rm base}$}&
 \colhead{Time} & 
 \colhead{Dev.} & 
 \colhead{Dev.}&
 \colhead{NLML} \\ 
 \colhead{} & \colhead{} & \multicolumn{2}{c}{($10^{-11}$\,\percmsqs)} &
 \colhead{(d)} & \colhead{} & \colhead{} & \colhead{}
}
\startdata
1ES 0806+524  & 4  & 2.1  & 0.6  & 0.14 & 0.81  & 0.10 &   4.47 \\
1ES 1218+304  & 20 & 4.5  & 1.6  & 0.55 & 1.25  & 0.08 &  29.68 \\
3C 66A 	      & 17 & 4.1  & 1.5  & 0.89 & 0.86  & 0.19 &  15.56 \\
Mrk 421       & 93 & 60.6 & 4.4  & 0.61 & 0.99  & 0.07 & 133.96 \\
Mrk 501       & 15 & 9.1  & 1.5  & 0.52 & 1.60  & 0.10 &  27.44 \\
W Comae       & 19 & 4.8  & 0.5  & 0.99 & 1.63  & 0.10 &  32.05 \\
\enddata

\tablecomments{Interpolated curves were generated using Gaussian
  likelihood, exact inference, a constant mean function and a squared
  exponential covariance function with isotropic distances. Maximum
  and baseline fluxes ($F_{\rm max}$, $F_{\rm base}$) are over
  $0.2\,{\rm TeV} < \egam < 30\,{\rm TeV}$. $F_{\rm base}$ is the
  value of the mean function and was estimated as the 
  mode of a Gaussian kernel density estimator, as applied to the
  complete set of observations for each source. The covariance
  function has two hyperparameters, the log of the correlation time
  and the log of the covariance noise. The likelihood function
  hyperparameter is the natural log of the standard deviation of the
  signal noise. The hyperparameters were optimized by minimizing the
  negative log marginal likelihood (NLML) with respect to the
  hyperparameters.}
\end{deluxetable*}


Blazars show variability on timescales as small as hours or even
minutes. This is reflected in the correlation lengths from the GPR,
which are all less than a day, implying flux measurements separated by
more than a day are effectively uncorrelated. The
correlation length for 1ES 0806+524 was significantly shorter than the
other blazars, though due to a small sample size, the value is not
well constrained. Better estimates of the correlation length would
require more extensive monitoring or historical data, as was used for
Mrk~421 and Mrk~501.
 
\begin{figure}
\includegraphics[width=\columnwidth]{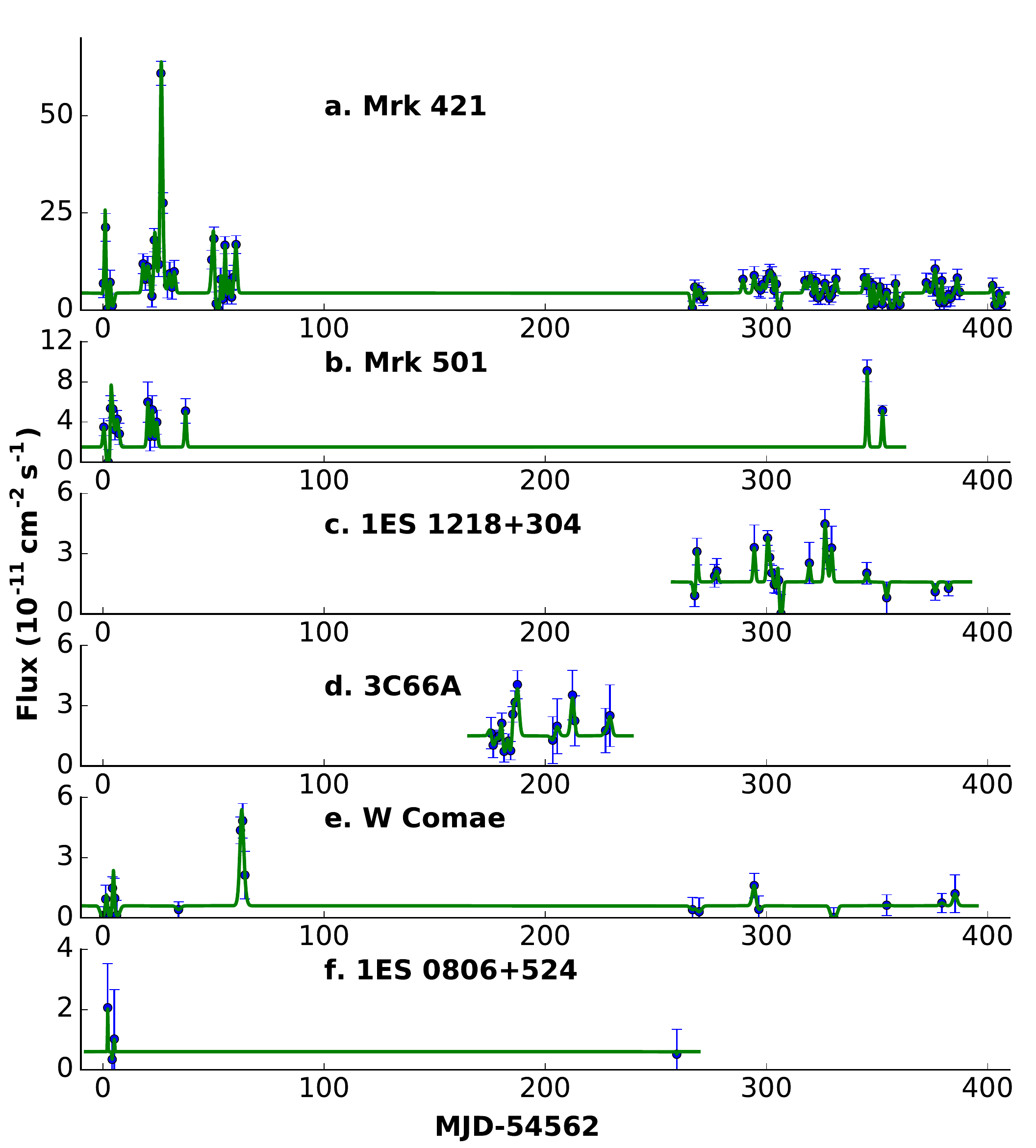}
\caption{\veritas\ TeV \gray\ light curves for six northern blazars
  subject to regular monitoring during the IC40 run, along with the
  maximum-likelihood interpolation resulting from our Gaussian process
  regression (Sec.~\ref{sub:lightcurves}). Time is in days since the
  start of IC40 (MJD 54562), aligned across all lightcurves.}
\label{fig:interpcurve}
\end{figure}

The resulting maximum-likelihood GPR-interpolated TeV \gray\ light
curves for our six target blazars, which we generate at one hour
resolution, are presented in Fig.~\ref{fig:interpcurve}. For both
Mrk~501 and 1ES 1218+304, there exist short periods where the
interpolated curve overshoots the measured flux points. This behavior
does not impact the selection of temporal windows of interest;
however, if the blazar emission did not actually increase during these
intervals, then it will lead to a slight overestimation of the
integrated TeV \gray\ fluence in these cases.


\subsection{Analysis Approach}
\label{sub:analysis}

Before using these interpolated light curves to define (statistically
optimal) temporal windows of interest for our search for an excess of
associated neutrinos, we need to understand the expected rate of
background atmospheric neutrinos. This requires choosing a region of
interest (ROI), that is, an acceptance radius (angle) around each blazar
position. \ice\ has established a 0.7\arcdeg\ uncertainty in track
event reconstruction \citep{icmoonshadow2013}. We treat this as the
standard deviation for a two-dimensional Gaussian distribution and
find that 99.5\% of neutrinos should be found within a
2.3\arcdeg\ radius of the reconstructed direction. We adopt this
2.3\arcdeg\ radius to define the ROI for each blazar. 

Selection of the temporal windows of interest begins with the
observation that the background atmospheric neutrinos are distributed
uniformly in time and near-uniformly on the sky
\citep{IC-40,icpoints2015}. The chief exception is the excess of
events near the horizon owing to atmospheric muons from the southern
hemisphere being (erroneously) reconstructed to low northern
declinations. Since none of the targeted blazars have declinations
$\delta_{2000} < +25\arcdeg$, we can safely ignore this effect.
Hence, the number of background neutrinos within the
2.3$^{\circ}$-radius ROI for any blazar is expected to follow Poisson
statistics, with an expected number of 0.023 neutrinos per day
($\approx$1 neutrino per ROI per 44 days). (Note that in the next
section we test this assumption using Monte Carlo simulations.)

Having generated interpolated light curves in TeV \grays\ for all six
monitored blazars, we use their variability to our advantage by
selecting the brightest emission periods for our search, maximizing
our sensitivity to flux-correlated neutrino emission in excess of the
atmospheric background. 

Considering the union of the six light curves at one hour resolution,
we sort flux measurements (irrespective of target blazar) from
brightest to faintest. This sorting yields a one-to-one mapping from
any given TeV flux threshold to a corresponding total exposure time
(with reduced thresholds implying greater integration times), along
with an associated total integrated fluence in TeV \grays. The total
exposure time, along with the angular size of our adopted ROI, implies
an expected number of background atmospheric neutrinos, which in turn
implies a minimum number of blazar-associated neutrinos which would
have to be present to yield a $>$3$\sigma$ detection assuming zero
detected atmospheric neutrinos. Our optimization is as follows: We
select the flux threshold that maximizes the ratio of the integrated
TeV fluence to the number of neutrinos needed for $>$3$\sigma$
detection.

\begin{figure}
\includegraphics[width=\columnwidth]{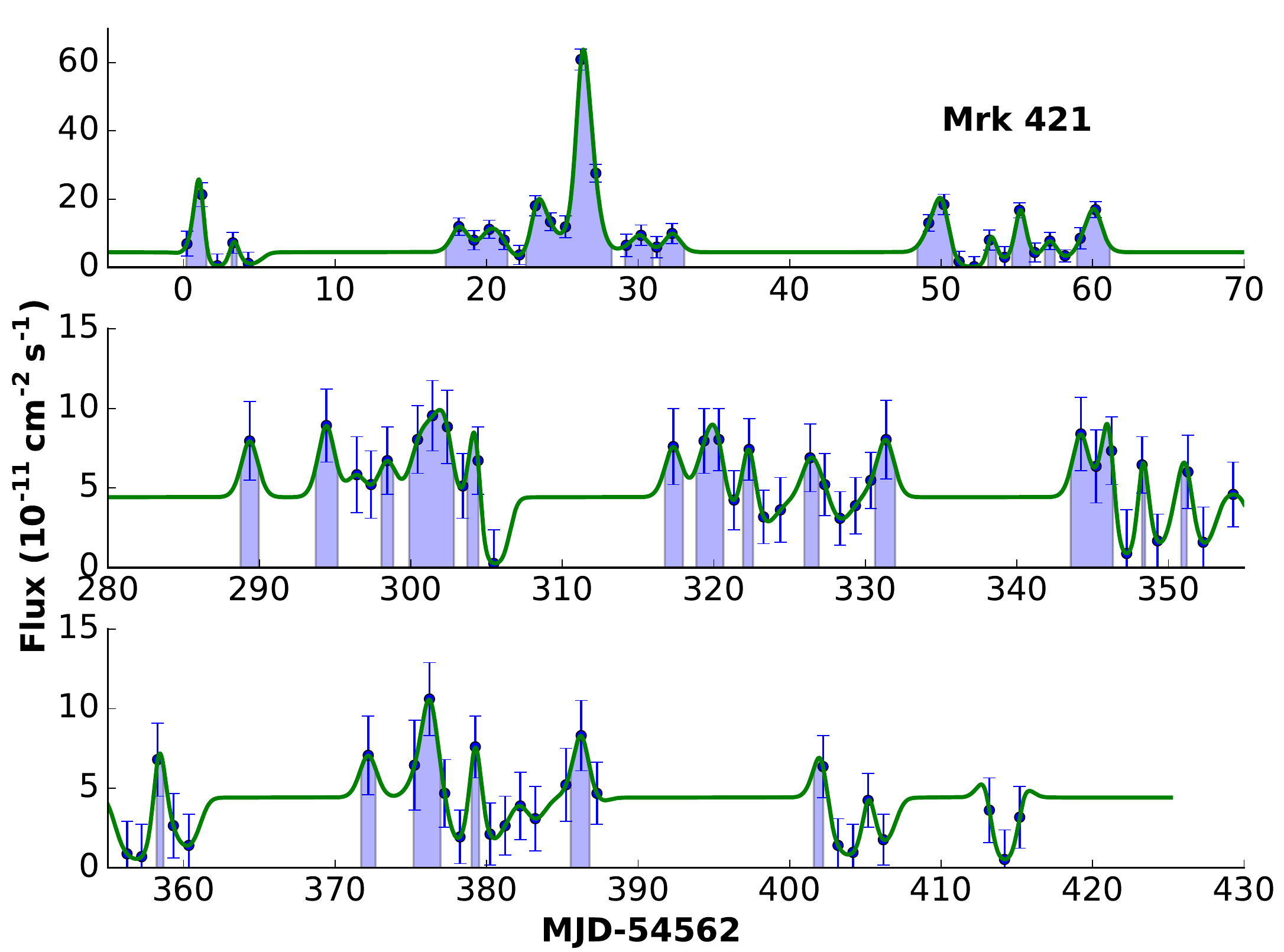}
\caption{Times of interest for Mrk~421 (shaded intervals). These
  times were selected in our initial optimization as the most
  sensitive search for associated neutrinos
  (Sec.~\ref{sub:analysis}). The selection includes 45.6 days with a
  total TeV \gray\ fluence of $4.1\times 10^{-4}$\,\phtpercm\ and
  yields an expected background of 1.03~neutrinos.}
\label{fig:msp}
\end{figure}

\begin{figure}
\includegraphics[width=\columnwidth]{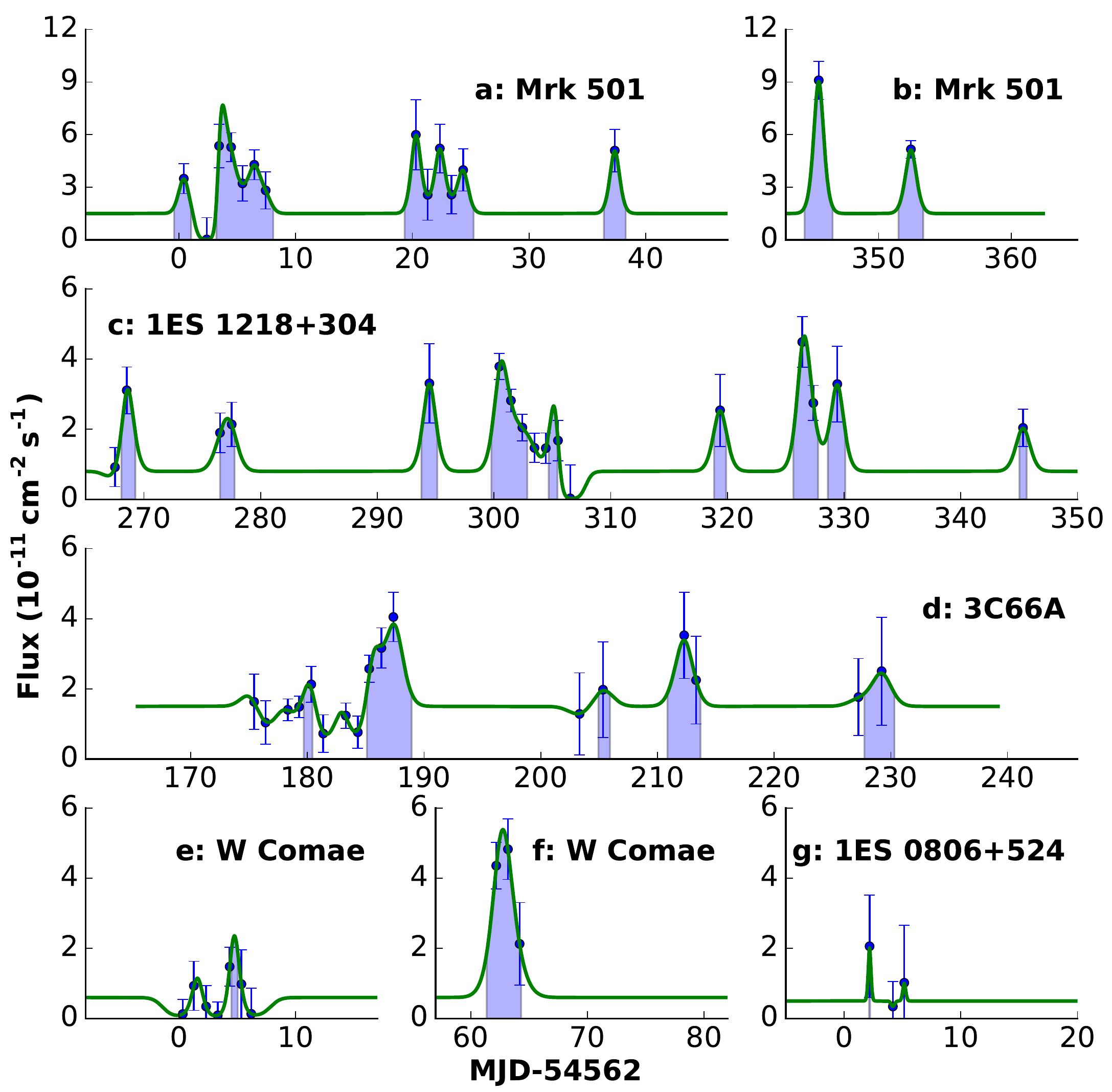}
\caption{Times of interest for Mrk~501, 1ES~1218+304, 3C~66A, W~Comae,
  and 1ES~0806+524 resulting from our Mrk~421-exclusive search (shaded
  intervals). These times were selected as the most sensitive search
  for associated neutrinos that excludes data from Mrk~421
  (Sec.~\ref{sub:analysis}). The selection includes 51.6 days with a
  total \gray\ fluence of $1.4\times 10^{-4}$\,\phtpercm\ and yields
  an expected background of 1.17~neutrinos.}
\label{fig:zsp}
\end{figure}

Our prior expectation was that this optimization would yield, for the
optimal flux threshold, multiple disjoint temporal acceptance windows
across the brightest emission periods of several distinct blazars. In
fact, however, when applied to the data for all six blazars, the
optimal flux threshold yields acceptance windows totalling
$\approx$46~days from Mrk~421 only, which is often the brightest
blazar in the sample. These temporal windows of interest are shown in
Fig.~\ref{fig:msp} and presented in Table~\ref{tab:flares}.
\added{The temporal window includes 11\% of the IC40 live time, and
  close to 18\% of the live time for Mrk~421, which is close to the
  3$\sigma$ duty cycle calculated in \citet{flareclass}.} Though this
is the optimal window as defined by our analysis, it completely omits
the five remaining blazars. So we choose to carry out two distinct
searches for excess neutrino emission: First, a Mrk~421-only analysis
(the optimal search, according to our original search criteria); and
second, a Mrk~421-exclusive analysis using the light curves from the
five other blazars. Optimization of the flux threshold for the second
analysis results in a total $\approx$52~day-long acceptance window
including some coverage of each of the five remaining blazars, as
shown in Fig.~\ref{fig:zsp} and presented in Table~\ref{tab:flares}.


\begin{deluxetable}{lrr}
\tabletypesize{\scriptsize}
\tablecolumns{3}
\tablewidth{1.5in}
\tablecaption{Blazar Times of Interest \label{tab:flares}}
\tablehead{
 \colhead{Name} &
 \colhead{$T_{\rm start}$} &
 \colhead{$T_{\rm stop}$} \\
 \colhead{} &
 \colhead{(d)} &
 \colhead{(d)}
}
\startdata
1ES 0806+524  & 2.138    & 2.221   \\
1ES 1218+304  & 268.058  & 269.266 \\
 ~            & 276.516  & 277.766 \\
 ~            & 293.766  & 295.141 \\
 ~            & 299.766  & 302.849 \\
 ~            & 304.683  & 305.433 \\
 ~            & 318.849  & 319.891 \\
 ~            & 325.641  & 327.766 \\
 ~            & 328.599  & 330.099 \\
 ~            & 345.016  & 345.641 \\
3C66A         & 179.687  & 180.437 \\
 ~            & 185.104  & 188.937 \\
 ~            & 204.979  & 205.937 \\
 ~            & 210.854  & 213.687 \\
 ~            & 227.729  & 230.312 \\
Mrk 421       & 0.116    & 1.532   \\
 ~            & 3.199    & 3.532   \\
 ~            & 17.324   & 21.366  \\
 ~            & 22.616   & 28.241  \\
 ~            & 29.157   & 30.949  \\
 ~            & 31.449   & 33.032  \\
 ~            & 48.449   & 50.741  \\
 ~            & 53.116   & 53.616  \\
 ~            & 54.699   & 55.866  \\
 ~            & 56.866   & 57.532  \\
 ~            & 58.949   & 61.157  \\
 ~            & 267.574  & 267.824 \\
 ~            & 288.741  & 289.991 \\
 ~            & 293.699  & 295.199 \\
 ~            & 298.074  & 298.824 \\
 ~            & 299.907  & 302.866 \\
 ~            & 303.741  & 304.449 \\
 ~            & 316.782  & 317.949 \\
 ~            & 318.824  & 320.657 \\
 ~            & 321.949  & 322.616 \\
 ~            & 325.991  & 326.907 \\
 ~            & 330.657  & 331.991 \\
 ~            & 343.574  & 346.324 \\
 ~            & 348.241  & 348.449 \\
 ~            & 350.866  & 351.199 \\
 ~            & 358.241  & 358.657 \\
 ~            & 371.741  & 372.657 \\
 ~            & 375.199  & 376.991 \\
 ~            & 379.032  & 379.532 \\
 ~            & 385.574  & 386.782 \\
 ~            & 401.616  & 402.199 \\
Mrk 501       & $-$0.961 & 0.830   \\
 ~            & 3.247    & 8.289   \\
 ~            & 18.705   & 25.789  \\
 ~            & 35.872   & 38.83   \\
 ~            & 343.872  & 347.122 \\
 ~            & 350.955  & 353.914 \\
W Comae       & 4.494    & 5.036   \\
 ~            & 61.411   & 64.327  \\
\enddata
\tablecomments{Start and stop times ($T_{\rm start}$, $T_{\rm stop}$)
  for the temporal windows of interest are reported in days since the
  start of the IC40 run (i.e., MJD$-$54562). Temporal windows for two
  distinct searches are reported, one using Mrk~421 data only
  (Mrk~421-only) and one using only data from other blazars
  (Mrk~421-exclusive); see text for details.} 
\end{deluxetable}


\subsection{Monte Carlo Simulations}
\label{sub:mcsims}

We carry out Monte Carlo (MC) simulations to confirm our understanding
of the atmospheric neutrino background. Each simulation begins by
shuffling the complete list of IC40 neutrino detections, associating
each original neutrino $\nu_i$ with another randomly-selected neutrino
$\nu_j$. In our MC scrambling, the neutrino $\nu_i$ retains its
original declination and receives the original arrival time of
neutrino $\nu_j$, with its new right ascension derived by adjusting
the original right ascension for the difference in local sidereal time
between the original and new arrival times. In this way, each neutrino
retains its original arrival direction with respect to the physical
IC40 array. Applying this procedure leads to a new scrambled dataset
that retains the original integrated distributions in time,
declination, and arrival direction, while scrambling neutrino
positions quasi-randomly in right ascension.

We produce 10,000 MC datasets and search for neutrinos arriving
during the predefined ROIs and temporal windows. We then use these
results to verify the expected arrival rate and assumption of Poisson
behavior for the background atmospheric neutrinos.

Our Mrk~421-only search uses a flux cutoff accepting each
lightcurve point brighter than $6.22\times 10^{-11}\percm\ \persec$,
accounting for 45.6 days in total. The integrated fluence during this
temporal window is $4.1\times 10^{-4}\percm$. From the
IC40 data, an average of 1.03 neutrinos should arrive within the
2.3$^{\circ}$-radius ROI during the 45.6 days. The 10,000 MC data sets
show a Poisson background with a mean of 1.03 and a
variance of 1.01. Thus, a $>$3$\sigma$ detection would require $\geq$5
neutrinos during the temporal acceptance window.

Our Mrk~421-exclusive search uses a flux cutoff of $ 1.88\times
10^{-11} \percm\ \persec$, with 51.6 days above this level, and an
integrated fluence of $ 1.4\times 10^{-4} \percm$. IC40 data predicts
the ROI should see an average of 1.17 neutrinos during the temporal
windows. The 10,000 scrambles demonstrated a Poisson background with a
mean and a variance of 1.10. Detection at $>$3$\sigma$ would hence
require $\geq$6 neutrinos.


\section{Results and Discussion}
\label{sec:discuss}

In order to derive physical constraints from our observations, we need
to derive expected IC40 neutrino detection rates from the observed
TeV \gray\ fluxes of the targeted blazars, using an appropriate
model. 

We adopt a standard lepto-hadronic model for this purpose. The
measured TeV photon fluence for each blazar can be expressed as the
integral of the differential spectrum:
\begin{equation}
   \phi_i = \int_{\varepsilon_{\rm th}}^{\varepsilon_2} \frac{N_0}{\egam^{2}}
   \left( \frac{\egam}{1\,{\rm TeV}}\right)^{2-\Gamma_{{\rm obs},i}}
   \, d\egam
   \label{eq:gamma}
\end{equation}
where $\egam$ is the \gray\ energy, $\varepsilon_{\rm th}$ is the
threshold energy for that blazar, $\varepsilon_2 = 30$\,TeV is the
upper energy limit of \veritas, $\Gamma_{{\rm obs},i}$ is the observed
power-law photon index for each blazar (Table~\ref{tab:blzinfo}), and
$N_0$ is a normalization constant. Once $N_0$ is known for each
blazar, the differential spectrum for neutrinos can be calculated. The
number of expected neutrinos is the integral of the neutrino spectrum
times the effective area of IC40:
\begin{equation}  
  N_{{\rm exp},i} =\int_{\varepsilon_1}^{\varepsilon_2} \frac{N_0}{4\,
    \enu^2} \left(\frac{\enu}{0.5\,{\rm TeV}}\right)^{2-\Gamma_{{\rm
        src},i}} A_{\rm IC40}(\enu) \, d\enu
  \label{eq:nu}
\end{equation}
where $N_0$ is a constant calculated in Eq.~\ref{eq:gamma}, $\enu$ is
the neutrino energy, $\varepsilon_1$ and $\varepsilon_2$ bound the
calculation over the IC40 effective area $A_{\rm IC40}(\epsilon_\nu)$
as presented in \citet{icarea}, and $\Gamma_{{\rm src},i}$ is the blazar spectral
index after correcting for \gray\ absorption by the extragalactic
background light using the models of \citet{ebl1} and \citet{eblinoue}
(Table~\ref{tab:blzinfo}). If $N_{{\rm exp},i}$ is larger than the
calculated upper limit, non-observation will constrain the fraction of
\gray s produced in lepto-hadronic interactions.

Our Mrk~421-only search yields no neutrinos during the temporal
acceptance windows, while $\ge$5 were required for a detection claim.
Our non-detection implies an upper limit of 1.27 (2.30 in total, minus
1.03 background) flux-associated neutrinos from Mrk~421 at
90\%-confidence. If the TeV $\gamma$-ray flux from Mrk~421 during our
identified periods of excess and flaring emission were entirely due to
hadronic interactions, this would give 0.79 neutrinos. As this is
smaller than the upper limit, we cannot constrain the hadronic process
for Mrk~421; the relevant figure of merit is $\nufom < 1.6$.

Our Mrk~421-exclusive search yields one neutrino during the joint
acceptance windows for the remaining five blazars; $\ge$6 neutrinos
were required in this case for $>$3$\sigma$ detection. The neutrino
arrives from the Mrk~501-specific ROI with an angular offset of
1.5$^{\circ}$ at MJD 54562.71 (time 0.71 in Figure \ref{fig:zsp} and
Table~\ref{tab:flares}). With one detected neutrino, we have an upper
limit of 2.79 (3.89 total minus 1.10 background) neutrinos at
90\%-confidence. Applying our hadronic interaction model to the
identified periods of excess and flaring emission from these blazars
leads to an expectation of 0.02 neutrinos over the joint acceptance
windows with $\nufom < 140$ as the figure of merit.  Note that while
the integrated \gray\ fluence from Mrk~421 was roughly three times
larger than the fluence from the other five blazars, its expected
neutrino fluence was nearly 35 times larger. This is due to the
hardness of Mrk~421's spectrum, which yields more neutrinos at higher
energies where \ice\ is more sensitive. The five other blazars
(excepting 3C 66A, which had a much smaller integrated fluence) had
softer spectra, yielding more neutrinos at lower energies where
\ice\ is less sensitive.

\added{A previously published IC40 analysis \citep{IC-40} for Mrk~421
  identified and observed two neutrinos, consistent with background,
  with theoretical neutrino flux calculations yielding a best fit of
  2.6 signal events, producing a differential flux limit of $ dN/dE
  \leq 11.71 \times 10^{-12}$\, $\rm TeV^{-1}$\, \percmsqs \ for muon
  neutrinos. Additional searches \citep{icfouryears} continue to show
  no correlated neutrinos.  Independent analysis of Mrk~421 using
  \gray\ data from Fermi \citep{nuexpect} showed that focusing on
  blazar flares may be optimal for discovery of correlated $>$PeV
  neutrinos.}

While our limits do not suffice to constrain our chosen hadronic
interaction model for these blazars, the existence of an additional
seven years' worth of \ice\ data, all taken with greater-volume
configurations of the detector (including five years' worth of data
using all 86 strings), is encouraging. This should amount to roughly
14 times the effective integration of one year of IC40 data and, in
the event of a nondetection, reduce the figure of merit by a factor of
${\approx}3.8$. Hence, unless the variability and flaring properties
of Mrk~421 and the remaining blazars during IC40 were in some way
unusual, we anticipate being able to constrain hadronic models for
Mrk~421 with data already in hand at both \ice\ and \veritas, with the
anticipated figure of merit (in the case of nondetection) being
$\nufom \simlt 0.4$. Similarly but less conclusively, we anticipate
being able to make a much more sensitive (although potentially, not
yet physically constraining) search for excess flux-associated
neutrinos from Mrk~501 and the other northern blazars, with $\nufom
\simlt 32$ as the anticipated figure of merit in the case of
non-detection.


\section{Conclusions}
\label{sec:conclude}

We defined a search for TeV \gray\ flux-correlated neutrinos from six
bright northern blazars. The search used publicly-available neutrino
data from the \ice\ 40-string ``IC40'' observing run and public TeV
$\gamma$-ray data from \veritas.  Interpolating the \veritas\ light
curves with a Gaussian process regression, we isolated temporal
windows of interest for two searches: a Mrk~421-only search and a
Mrk~421-exclusive search. We confirmed the Poisson behavior of the
near-isotropic background with a Monte Carlo simulation using
scrambled neutrino datasets.

Our Mrk~421-only search found zero neutrinos compared to a background
expectation of 1.03 neutrinos and a requirement of $\ge$5 neutrinos
for a $>$3$\sigma$ detection claim, while our Mrk~421-exclusive search
found one neutrino compared to a background expectation of 1.1
neutrinos and a requirement of $\ge$6 neutrinos for a $>$3$\sigma$
detection claim.

Both findings are consistent with background expectations, yet they
are also consistent with expectations from hadronic blazar
models. These non-detections place upper limits of 1.27 neutrinos for
Mrk~421 (with a figure of merit $\nufom < 1.6$) and 2.79 neutrinos for
the five other blazars (figure of merit $\nufom < 140$).  These limits
are not strong enough to place constraints on the hadronic process for
any of the blazars, though the figure of merit for Mrk~421 is close to
being physically constraining.  However, this search was limited to
only the IC40 dataset. The methods developed in
Sec.~\ref{sub:analysis} work equally well for any additional existing
data. An interesting future project would be to extend this search to
more recent public datasets.

Our analysis showed that Mrk~421 dominates the TeV $\gamma$-ray flux
from the northern blazars. Mrk~421 is currently monitored in the TeV
on a daily or near-daily basis by the High-Altitude Water Cherenkov
(HAWC; \citealt{hawcdesign}) and First Cherenkov Telescope using a
G-APD Camera for TeV Gamma-ray Astronomy (FACT; \citealt{factdesign})
facilities. Since \ice\ has been operating at full strength since
2011, it may be fruitful to perform a flux-correlated search on
Mrk~421, as even a non-detection will reduce the figure of merit,
potentially to a physically constraining value.

As mentioned in Sec.~\ref{sec:intro}, TeV orphan flares are a
particular candidate for the dominance of hadronic acceleration
processes. Using hard \xray\ data from the Swift BAT all-sky monitor
\citep{batmonitor} in tandem with HAWC, FACT, and VERITAS data, TeV
\gray\ orphan flares could be isolated from other periods of high TeV
flux and used for a distinct, focused seach for associated neutrinos. 


\vspace{0.5\baselineskip}

\acknowledgements

The authors acknowledge support from the Eberly College of Science,
the Penn State Institute for Gravitation and the Cosmos, and the
partner collaborations of the Astrophysical Multimessenger Observatory
Network (AMON). The authors would like to thank M.~Errando,
E.~Pueschel, and \mbox{S.~O'Brien} from the VERITAS collaboration for
their efforts in preparing and publishing the blazar light curves. The
authors acknowledge statistical consulting support from
Prof.\ B.~Shaby of the Penn State Department of Statistics. This work
was supported in part by the National Science Foundation under Grant
Number \mbox{PHY-1412633}.

\facilities{IceCube, VERITAS, Whipple Telescope}

\software{SciPy \citep{scipy}, NumPy \citep{numpy}, pyGPs \citep{pygps}, matplotlib \citep{matplotlib}}


\bibliographystyle{apj_8}
\bibliography{ictev}

\begin{thebibliography}{73}
\expandafter\ifx\csname natexlab\endcsname\relax\def\natexlab#1{#1}\fi

\bibitem[{{Aartsen} {et~al.}(2013{\natexlab{a}}){Aartsen}, {Abbasi}, {Abdou},
  {Ackermann}, {Adams}, {Aguilar}, {Ahlers}, {Altmann}, {Auffenberg}, {Bai}, \&
  et~al.}]{Ic3+13-sci}
{Aartsen}, M.~G. {et al.}\  2013{\natexlab{a}}, Science, 342, 1

\bibitem[{{Aartsen} {et~al.}(2013{\natexlab{b}}){Aartsen}, {Abbasi}, {Abdou},
  {Ackermann}, {Adams}, {Aguilar}, {Ahlers}, {Altmann}, {Auffenberg}, {Bai}, \&
  et~al.}]{ic-pev}
--- 2013{\natexlab{b}}, Physical Review Letters, 111, 021103

\bibitem[{{Aartsen} {et~al.}(2013{\natexlab{c}}){Aartsen}, {Abbasi}, {Abdou},
  {Ackermann}, {Adams}, {Aguilar}, {Ahlers}, {Altmann}, {Auffenberg}, \&
  et~al.}]{icmoonshadow2013}
--- 2013{\natexlab{c}}, ArXiv.org, 1305.6811

\bibitem[{{Aartsen} {et~al.}(2015){Aartsen}, {Ackermann}, {Adams}, {Aguilar},
  {Ahlers}, {Ahrens}, {Altmann}, {Anderson}, {Archinger}, {Arguelles}, \&
  et~al.}]{icpoints2015}
--- 2015, \apj, 807, 46

\bibitem[{{Aartsen} {et~al.}(2014{\natexlab{a}}){Aartsen}, {Ackermann},
  {Adams}, {Aguilar}, {Ahlers}, {Ahrens}, {Altmann}, {Anderson}, {Arguelles},
  {Arlen}, \& et~al.}]{Ic3+14-tevpev}
--- 2014{\natexlab{a}}, Physical Review Letters, 113, 101101

\bibitem[{{Aartsen} {et~al.}(2014{\natexlab{b}}){Aartsen}, {Ackermann},
  {Adams}, {Aguilar}, {Ahlers}, {Ahrens}, {Altmann}, {Anderson}, {Arguelles},
  {Arlen}, \& et~al.}]{icfouryears}
--- 2014{\natexlab{b}}, \apj, 796, 109

\bibitem[{{Aartsen} {et~al.}(2014{\natexlab{c}}){Aartsen}, {Ackermann},
  {Adams}, {Aguilar}, {Ahlers}, {Ahrens}, {Altmann}, {Anderson}, {Arguelles},
  \& et~al.}]{icgrb14}
--- 2014{\natexlab{c}}, ArXiv.org, 1412.6510

\bibitem[{{Abbasi} {et~al.}(2012){Abbasi}, {Abdou}, {Abu-Zayyad}, {Ackermann},
  {Adams}, {Aguilar}, {Ahlers}, {Altmann}, {Andeen}, {Auffenberg}, \&
  et~al.}]{icgrb12}
{Abbasi}, R. {et al.}\  2012, \nat, 484, 351

\bibitem[{{Abbasi} {et~al.}(2011{\natexlab{a}}){Abbasi}, {Abdou}, {Abu-Zayyad},
  {Adams}, {Aguilar}, {Ahlers}, {Altmann}, {Andeen}, {Auffenberg}, {Bai}, \&
  et~al.}]{icarea}
--- 2011{\natexlab{a}}, \prd, 84, 082001

\bibitem[{{Abbasi} {et~al.}(2011{\natexlab{b}}){Abbasi}, {Abdou}, {Abu-Zayyad},
  {Adams}, {Aguilar}, {Ahlers}, {Andeen}, {Auffenberg}, {Bai}, {Baker}, \&
  et~al.}]{IC-40}
--- 2011{\natexlab{b}}, \apj, 732, 18

\bibitem[{{Abdo} {et~al.}(2012){Abdo}, {Abeysekara}, {Allen}, {Aune}, {Berley},
  {Bonamente}, {Christopher}, {DeYoung}, {Dingus}, {Ellsworth},
  {Galbraith-Frew}, {Gonzalez}, {Goodman}, {Hoffman}, {H{\"u}ntemeyer}, {Hui},
  {Kolterman}, {Linnemann}, {McEnery}, {Mincer}, {Morgan}, {Nemethy}, {Pretz},
  {Ryan}, {Saz Parkinson}, {Shoup}, {Sinnis}, {Smith}, {Vasileiou}, {Walker},
  {Williams}, \& {Yodh}}]{milagro12_cygnus}
{Abdo}, A.~A. {et al.}\  2012, \apj, 753, 159

\bibitem[{{Acciari} {et~al.}(2014){Acciari}, {Arlen}, {Aune}, {Benbow}, {Bird},
  {Bouvier}, {Bradbury}, {Buckley}, {Bugaev}, {de la Calle Perez},
  {Carter-Lewis}, {Cesarini}, {Ciupik}, {Collins-Hughes}, {Connolly}, {Cui},
  {Duke}, {Dumm}, {Falcone}, {Federici}, {Fegan}, {Fegan}, {Finley},
  {Finnegan}, {Fortson}, {Gaidos}, {Galante}, {Gall}, {Gibbs}, {Gillanders},
  {Griffin}, {Grube}, {Gyuk}, {Hanna}, {Horan}, {Humensky}, {Kaaret},
  {Kertzman}, {Khassen}, {Kieda}, {Krawczynski}, {Krennrich}, {Lang},
  {McEnery}, {Madhavan}, {Moriarty}, {Nelson}, {O'Faol{\'a}in de Bhr{\'o}ithe},
  {Ong}, {Orr}, {Otte}, {Perkins}, {Petry}, {Pichel}, {Pohl}, {Quinn}, {Ragan},
  {Reynolds}, {Roache}, {Rovero}, {Schroedter}, {Sembroski}, {Smith},
  {Telezhinsky}, {Theiling}, {Toner}, {Tyler}, {Varlotta}, {Vivier}, {Wakely},
  {Ward}, {Weekes}, {Weinstein}, {Welsing}, {Williams}, \&
  {Wissel}}]{14ymrk421}
{Acciari}, V.~A. {et al.}\  2014, Astroparticle Physics, 54, 1

\bibitem[{{Achterberg} {et~al.}(2006){Achterberg}, {Ackermann}, {Adams},
  {Ahrens}, {Andeen}, {Atlee}, {Baccus}, {Bahcall}, {Bai}, \&
  et~al.}]{icfacility}
{Achterberg}, A. {et al.}\  2006, Astroparticle Physics, 26, 155

\bibitem[{{Adri{\'a}n-Mart{\'{\i}}nez}
  {et~al.}(2013){Adri{\'a}n-Mart{\'{\i}}nez}, {Albert}, {Samarai}, {Andr{\'e}},
  {Anghinolfi}, {Anton}, {Anvar}, {Ardid}, {Astraatmadja}, {Aubert}, {Baret},
  {Barrios-Marti}, {Basa}, {Bertin}, {Biagi}, {Bigongiari}, {Bogazzi},
  {Bouhou}, {Bouwhuis}, {Brunner}, {Busto}, {Capone}, {Caramete},
  {C{\^a}rloganu}, {Carr}, {Cecchini}, {Charif}, {Charvis}, {Chiarusi},
  {Circella}, {Classen}, {Coniglione}, {Core}, {Costantini}, {Coyle},
  {Creusot}, {Curtil}, {De Bonis}, {Dekeyser}, {Deschamps}, {Distefano},
  {Donzaud}, {Dornic}, {Dorosti}, {Drouhin}, {Dumas}, {Eberl}, {Emanuele},
  {Enzenh{\"o}fer}, {Ernenwein}, {Escoffier}, {Fehn}, {Fermani}, {Flaminio},
  {Folger}, {Fritsch}, {Fusco}, {Galat{\`a}}, {Gay}, {Gei{\ss}els{\"o}der},
  {Geyer}, {Giacomelli}, {Giordano}, {Gleixner}, {G{\'o}mez-Gonz{\'a}lez},
  {Graf}, {Guillard}, {van Haren}, {Heijboer}, {Hello}, {Hern{\'a}ndez-Rey},
  {Herold}, {H{\"o}{\ss}l}, {James}, {de Jong}, {Kadler}, {Kalekin}, {Kappes},
  {Katz}, {Kooijman}, {Kouchner}, {Kreykenbohm}, {Kulikovskiy}, {Lahmann},
  {Lambard}, {Lambard}, {Larosa}, {Lef{\`e}vre}, {Leonora}, {Lo Presti},
  {Loehner}, {Loucatos}, {Louis}, {Mangano}, {Marcelin}, {Margiotta},
  {Mart{\'{\i}}nez-Mora}, {Martini}, {Michael}, {Montaruli}, {Morganti},
  {M{\"u}ller}, {Neff}, {Nezri}, {Palioselitis}, {P{\u a}v{\u a}la{\c s}},
  {Perrina}, {Piattelli}, {Popa}, {Pradier}, {Racca}, {Riccobene}, {Richter},
  {Rivi{\`e}re}, {Robert}, {Roensch}, {Rostovtsev}, {Samtleben}, {Sanguineti},
  {Schmid}, {Schnabel}, {Schulte}, {Sch{\"u}ssler}, {Seitz}, {Shanidze},
  {Sieger}, {Simeone}, {Spies}, {Spurio}, {Steijger}, {Stolarczyk},
  {S{\'a}nchez-Losa}, {Taiuti}, {Tamburini}, {Tayalati}, {Trovato}, {Vallage},
  {Vall{\'e}e}, {Van Elewyck}, {Vernin}, {Visser}, {Wagner}, {Wilms}, {de
  Wolf}, {Yatkin}, {Yepes}, {Zornoza}, {Z{\'u}{\~n}iga}, \&
  {Baerwald}}]{antares_grb13}
{Adri{\'a}n-Mart{\'{\i}}nez}, S. {et al.}\  2013, \aap, 559, A9

\bibitem[{{Aharonian}(2000)}]{ahar00}
{Aharonian}, F.~A. 2000, \na, 5, 377

\bibitem[{{Ahlers} \& {Murase}(2014)}]{2014PhRvD..90b3010A}
{Ahlers}, M. \& {Murase}, K. 2014, \prd, 90, 023010

\bibitem[{{Anchordoqui} {et~al.}(2014{\natexlab{a}}){Anchordoqui}, {Barger},
  {Cholis}, {Goldberg}, {Hooper}, {Kusenko}, {Learned}, {Marfatia}, {Pakvasa},
  {Paul}, \& {Weiler}}]{abc+14}
{Anchordoqui}, L.~A. {et al.}\  2014{\natexlab{a}}, Journal of High Energy
  Astrophysics, 1, 1

\bibitem[{{Anchordoqui} {et~al.}(2014{\natexlab{b}}){Anchordoqui}, {Goldberg},
  {Paul}, {da Silva}, \& {Vlcek}}]{AGPestimate2014}
{Anchordoqui}, L.~A., {Goldberg}, H., {Paul}, T.~C., {da Silva}, L.~H.~M., \&
  {Vlcek}, B.~J. 2014{\natexlab{b}}, \prd, 90, 123010

\bibitem[{{Anderhub} {et~al.}(2010){Anderhub}, {Backes}, {Biland}, {Boller},
  {Braun}, {Bretz}, {Commichau}, {Commichau}, {Domke}, {Dorner}, {Gendotti},
  {Grimm}, {von Gunten}, {Hildebrand}, {Horisberger}, {K{\"o}hne},
  {Kr{\"a}henb{\"u}hl}, {Kranich}, {Krumm}, {Lorenz}, {Lustermann}, {Mannheim},
  {Neise}, {Pauss}, {Renker}, {Rhode}, {Rissi}, {Ribordy}, {R{\"o}ser},
  {Stark}, {Stucki}, {Tibolla}, {Viertel}, {Vogler}, {Warda}, \&
  {Weitzel}}]{factdesign}
{Anderhub}, H. {et al.}\  2010, ArXiv.org, 1010.2397

\bibitem[{{Atoyan} \& {Dermer}(2001)}]{ad01}
{Atoyan}, A. \& {Dermer}, C.~D. 2001, Physical Review Letters, 87, 221102

\bibitem[{{Bahcall} \& {Waxman}(2001)}]{bw01}
{Bahcall}, J. \& {Waxman}, E. 2001, \prd, 64, 023002

\bibitem[{{B{\"o}ttcher}(2005)}]{blottcherorphan05}
{B{\"o}ttcher}, M. 2005, \apj, 621, 176

\bibitem[{{B{\"o}ttcher}(2007)}]{blazar-model2}
--- 2007, \apss, 309, 95

\bibitem[{{Budnik} {et~al.}(2008){Budnik}, {Katz}, {MacFadyen}, \&
  {Waxman}}]{Budnik+08hn}
{Budnik}, R., {Katz}, B., {MacFadyen}, A., \& {Waxman}, E. 2008, \apj, 673, 928

\bibitem[{{Bustamante} {et~al.}(2015){Bustamante}, {Baerwald}, {Murase}, \&
  {Winter}}]{bbm+15}
{Bustamante}, M., {Baerwald}, P., {Murase}, K., \& {Winter}, W. 2015, Nature
  Communications, 6, 6783

\bibitem[{{Cerruti} {et~al.}(2015){Cerruti}, {Zech}, {Boisson}, \&
  {Inoue}}]{czb+15}
{Cerruti}, M., {Zech}, A., {Boisson}, C., \& {Inoue}, S. 2015, \mnras, 448, 910

\bibitem[{{Crocker} \& {Aharonian}(2011)}]{cabubble2011}
{Crocker}, R.~M. \& {Aharonian}, F. 2011, Physical Review Letters, 106, 101102

\bibitem[{{Crocker} {et~al.}(2014){Crocker}, {Bicknell}, {Carretti}, {Hill}, \&
  {Sutherland}}]{cbcfermibubble2014}
{Crocker}, R.~M., {Bicknell}, G.~V., {Carretti}, E., {Hill}, A.~S., \&
  {Sutherland}, R.~S. 2014, \apjl, 791, L20

\bibitem[{{Dermer} {et~al.}(2012){Dermer}, {Murase}, \& {Takami}}]{dmt12}
{Dermer}, C.~D., {Murase}, K., \& {Takami}, H. 2012, \apj, 755, 147

\bibitem[{{DeYoung} \& {HAWC Collaboration}(2012)}]{hawcdesign}
{DeYoung}, T. \& {HAWC Collaboration} 2012, Nuclear Instruments and Methods in
  Physics Research A, 692, 72

\bibitem[{{Essey} {et~al.}(2010){Essey}, {Kalashev}, {Kusenko}, \&
  {Beacom}}]{ekk+10}
{Essey}, W., {Kalashev}, O.~E., {Kusenko}, A., \& {Beacom}, J.~F. 2010,
  Physical Review Letters, 104, 141102

\bibitem[{{Finke} {et~al.}(2010){Finke}, {Razzaque}, \& {Dermer}}]{ebl1}
{Finke}, J.~D., {Razzaque}, S., \& {Dermer}, C.~D. 2010, \apj, 712, 238

\bibitem[{{Fox} {et~al.}(2013){Fox}, {Kashiyama}, \&
  {M{\'e}szar{\'o}s}}]{fkm13}
{Fox}, D.~B., {Kashiyama}, K., \& {M{\'e}szar{\'o}s}, P. 2013, \apj, 774, 74

\bibitem[{{Gaisser} {et~al.}(1995){Gaisser}, {Halzen}, \& {Stanev}}]{ghsrev95}
{Gaisser}, T.~K., {Halzen}, F., \& {Stanev}, T. 1995, \physrep, 258, 173

\bibitem[{{Gao} {et~al.}(2013){Gao}, {Kashiyama}, \&
  {M{\'e}sz{\'a}ros}}]{gkm13}
{Gao}, S., {Kashiyama}, K., \& {M{\'e}sz{\'a}ros}, P. 2013, \apjl, 772, L4

\bibitem[{{Halzen} \& {Hooper}(2005)}]{hh05}
{Halzen}, F. \& {Hooper}, D. 2005, Astroparticle Physics, 23, 537

\bibitem[{{Halzen} \& {Zas}(1997)}]{hz97}
{Halzen}, F. \& {Zas}, E. 1997, \apj, 488, 669

\bibitem[{{Holder} {et~al.}(2006){Holder}, {Atkins}, {Badran}, {Blaylock},
  {Bradbury}, {Buckley}, {Byrum}, {Carter-Lewis}, {Celik}, {Chow}, {Cogan},
  {Cui}, {Daniel}, {de la Calle Perez}, {Dowdall}, {Dowkontt}, {Duke},
  {Falcone}, {Fegan}, {Finley}, {Fortin}, {Fortson}, {Gibbs}, {Gillanders},
  {Glidewell}, {Grube}, {Gutierrez}, {Gyuk}, {Hall}, {Hanna}, {Hays}, {Horan},
  {Hughes}, {Humensky}, {Imran}, {Jung}, {Kaaret}, {Kenny}, {Kieda}, {Kildea},
  {Knapp}, {Krawczynski}, {Krennrich}, {Lang}, {LeBohec}, {Linton}, {Little},
  {Maier}, {Manseri}, {Milovanovic}, {Moriarty}, {Mukherjee}, {Ogden}, {Ong},
  {Petry}, {Perkins}, {Pizlo}, {Pohl}, {Quinn}, {Ragan}, {Reynolds}, {Roache},
  {Rose}, {Schroedter}, {Sembroski}, {Sleege}, {Steele}, {Swordy}, {Syson},
  {Toner}, {Valcarcel}, {Vassiliev}, {Wakely}, {Weekes}, {White}, {Williams},
  \& {Wagner}}]{veritas2008}
{Holder}, J. {et al.}\  2006, Astroparticle Physics, 25, 391

\bibitem[{Hunter(2007)}]{matplotlib}
  Hunter, J.~D. 2007, Computing In Science \& Engineering, 9, 90

\bibitem[{{Inoue} {et~al.}(2013){Inoue}, {Inoue}, {Kobayashi}, {Makiya},
  {Niino}, \& {Totani}}]{eblinoue}
{Inoue}, Y., {Inoue}, S., {Kobayashi}, M.~A.~R., {Makiya}, R., {Niino}, Y., \&
  {Totani}, T. 2013, \apj, 768, 197

\bibitem[{Jones {et~al.}(2001--)Jones, Oliphant, Peterson, {et~al.}}]{scipy}
Jones, E., Oliphant, T., Peterson, P., {et~al.} 2001--, {SciPy}: Open source
  scientific tools for {Python}, [Online; accessed 2016-07-26]

\bibitem[{{Kadler} {et~al.}(2016){Kadler}, {Krau{\ss}}, {Mannheim}, {Ojha},
  {M{\"u}ller}, {Schulz}, {Anton}, {Baumgartner}, {Beuchert}, {Buson},
  {Carpenter}, {Eberl}, {Edwards}, {Eisenacher Glawion}, {Els{\"a}sser},
  {Gehrels}, {Gr{\"a}fe}, {Hase}, {Horiuchi}, {James}, {Kappes}, {Kappes},
  {Katz}, {Kreikenbohm}, {Kreter}, {Kreykenbohm}, {Langejahn}, {Leiter},
  {Litzinger}, {Longo}, {Lovell}, {McEnery}, {Phillips}, {Pl{\"o}tz}, {Quick},
  {Ros}, {Stecker}, {Steinbring}, {Stevens}, {Thompson}, {Tr{\"u}stedt},
  {Tzioumis}, {Wilms}, \& {Zensus}}]{pev2016}
{Kadler}, M. {et al.}\  2016, ArXiv.org, 1602.02012

\bibitem[{{Kildea} {et~al.}(2007){Kildea}, {Atkins}, {Badran}, {Blaylock},
  {Bond}, {Bradbury}, {Buckley}, {Carter-Lewis}, {Celik}, {Chow}, {Cui},
  {Cogan}, {Daniel}, {de la Calle Perez}, {Dowdall}, {Duke}, {Falcone},
  {Fegan}, {Fegan}, {Finley}, {Fortson}, {Gall}, {Gillanders}, {Grube},
  {Gutierrez}, {Hall}, {Hall}, {Holder}, {Horan}, {Hughes}, {Jordan}, {Jung},
  {Kenny}, {Kertzman}, {Knapp}, {Konopelko}, {Kosack}, {Krawczynski},
  {Krennrich}, {Lang}, {LeBohec}, {Lloyd-Evans}, {Millis}, {Moriarty}, {Nagai},
  {Ogden}, {Ong}, {Perkins}, {Petry}, {Pizlo}, {Pohl}, {Quinn}, {Quinn},
  {Rebillot}, {Rose}, {Schroedter}, {Sembroski}, {Smith}, {Syson}, {Toner},
  {Valcarcel}, {Vassiliev}, {Wakely}, {Weekes}, \& {White}}]{whip2007}
{Kildea}, J. {et al.}\  2007, Astroparticle Physics, 28, 182

\bibitem[{{Krawczynski} {et~al.}(2004){Krawczynski}, {Hughes}, {Horan},
  {Aharonian}, {Aller}, {Aller}, {Boltwood}, {Buckley}, {Coppi}, {Fossati},
  {G{\"o}tting}, {Holder}, {Horns}, {Kurtanidze}, {Marscher}, {Nikolashvili},
  {Remillard}, {Sadun}, \& {Schr{\"o}der}}]{orphanmulti}
{Krawczynski}, H. {et al.}\  2004, \apj, 601, 151

\bibitem[{{Krimm} {et~al.}(2013){Krimm}, {Holland}, {Corbet}, {Pearlman},
  {Romano}, {Kennea}, {Bloom}, {Barthelmy}, {Baumgartner}, {Cummings},
  {Gehrels}, {Lien}, {Markwardt}, {Palmer}, {Sakamoto}, {Stamatikos}, \&
  {Ukwatta}}]{batmonitor}
{Krimm}, H.~A. {et al.}\  2013, \apjs, 209, 14

\bibitem[{{Laha} {et~al.}(2013){Laha}, {Beacom}, {Dasgupta}, {Horiuchi}, \&
  {Murase}}]{lbd+13}
{Laha}, R., {Beacom}, J.~F., {Dasgupta}, B., {Horiuchi}, S., \& {Murase}, K.
  2013, \prd, 88, 043009

\bibitem[{{Loeb} \& {Waxman}(2006)}]{lwstarburst2006}
{Loeb}, A. \& {Waxman}, E. 2006, \jcap, 5, 3

\bibitem[{{Lunardini} {et~al.}(2014){Lunardini}, {Razzaque}, {Theodoseau}, \&
  {Yang}}]{lrtbubble2014}
{Lunardini}, C., {Razzaque}, S., {Theodoseau}, K.~T., \& {Yang}, L. 2014, \prd,
  90, 023016

\bibitem[{{Mannheim}(1993)}]{mannheim93}
{Mannheim}, K. 1993, \prd, 48, 2408

\bibitem[{{Maraschi} {et~al.}(1992){Maraschi}, {Ghisellini}, \&
  {Celotti}}]{mgcleptonic92}
{Maraschi}, L., {Ghisellini}, G., \& {Celotti}, A. 1992, \apjl, 397, L5

\bibitem[{{M{\'e}sz{\'a}ros}(2014)}]{meszarosgrb2014}
{M{\'e}sz{\'a}ros}, P. 2014, Nuclear Physics B Proceedings Supplements, 256,
  241

\bibitem[{{M{\'e}sz{\'a}ros} \& {Waxman}(2001)}]{mw01}
{M{\'e}sz{\'a}ros}, P. \& {Waxman}, E. 2001, Physical Review Letters, 87,
  171102

\bibitem[{{Murase}(2014)}]{murase14}
{Murase}, K. 2014, ArXiv.org, 1410.3680

\bibitem[{{Murase} {et~al.}(2013){Murase}, {Ahlers}, \& {Lacki}}]{mal13}
{Murase}, K., {Ahlers}, M., \& {Lacki}, B.~C. 2013, \prd, 88, 121301

\bibitem[{{Murase} {et~al.}(2012){Murase}, {Dermer}, {Takami}, \&
  {Migliori}}]{mdt+12}
{Murase}, K., {Dermer}, C.~D., {Takami}, H., \& {Migliori}, G. 2012, \apj, 749,
  63

\bibitem[{{Murase} {et~al.}(2008){Murase}, {Inoue}, \& {Nagataki}}]{kin08}
{Murase}, K., {Inoue}, S., \& {Nagataki}, S. 2008, \apjl, 689, L105

\bibitem[{{Murase} {et~al.}(2014){Murase}, {Inoue}, \&
  {Dermer}}]{muraseagn2014}
{Murase}, K., {Inoue}, Y., \& {Dermer}, C.~D. 2014, \prd, 90, 023007

\bibitem[{{Murase} \& {Ioka}(2013)}]{mi13}
{Murase}, K. \& {Ioka}, K. 2013, Physical Review Letters, 111, 121102

\bibitem[{{Neumann} {et~al.}(2014){Neumann}, {Huang}, {Marthaler}, \&
  {Kersting}}]{pygps}
{Neumann}, M., {Huang}, S., {Marthaler}, D., \& {Kersting}, K. 2014, {pyGPS}: A
  Package for Gaussian Processes, [Online; accessed 2016-07-26]

\bibitem[{{Petropoulou} {et~al.}(2016){Petropoulou}, {Coenders}, \&
  {Dimitrakoudis}}]{nuexpect}
{Petropoulou}, M., {Coenders}, S., \& {Dimitrakoudis}, S. 2016, Astroparticle
  Physics, 80, 115

\bibitem[{{Rasmussen} \& {Williams}(2006)}]{gpml}
{Rasmussen}, C.~E. \& {Williams}, C.~K.~I. 2006, Gaussian processes for machine
  learning (Cambridge, Mass: MIT Press)

\bibitem[{{Reimer} {et~al.}(2005){Reimer}, {B{\"o}ttcher}, \&
  {Postnikov}}]{orphan-flare}
{Reimer}, A., {B{\"o}ttcher}, M., \& {Postnikov}, S. 2005, \apj, 630, 186

\bibitem[{{Resconi} {et~al.}(2009){Resconi}, {Franco}, {Gross}, {Costamante},
  \& {Flaccomio}}]{flareclass}
{Resconi}, E., {Franco}, D., {Gross}, A., {Costamante}, L., \& {Flaccomio}, E.
  2009, \aap, 502, 499

\bibitem[{{Sahu} {et~al.}(2013){Sahu}, {Oliveros}, \&
  {Sanabria}}]{sosorphan2013}
{Sahu}, S., {Oliveros}, A.~F.~O., \& {Sanabria}, J.~C. 2013, \prd, 87, 103015

\bibitem[{{Senno} {et~al.}(2015){Senno}, {M{\'e}sz{\'a}ros}, {Murase},
  {Baerwald}, \& {Rees}}]{smm+15}
{Senno}, N., {M{\'e}sz{\'a}ros}, P., {Murase}, K., {Baerwald}, P., \& {Rees},
  M.~J. 2015, \apj, 806, 24

\bibitem[{{Stecker} {et~al.}(1991){Stecker}, {Done}, {Salamon}, \&
  {Sommers}}]{steckeragn1991}
{Stecker}, F.~W., {Done}, C., {Salamon}, M.~H., \& {Sommers}, P. 1991, Physical
  Review Letters, 66, 2697

\bibitem[{{Tamborra} {et~al.}(2014){Tamborra}, {Ando}, \&
  {Murase}}]{tamstarform2014}
{Tamborra}, I., {Ando}, S., \& {Murase}, K. 2014, \jcap, 9, 43

\bibitem[{{Tavecchio} {et~al.}(2011){Tavecchio}, {Becerra-Gonzalez},
  {Ghisellini}, {Stamerra}, {Bonnoli}, {Foschini}, \& {Maraschi}}]{tbg+11}
{Tavecchio}, F., {Becerra-Gonzalez}, J., {Ghisellini}, G., {Stamerra}, A.,
  {Bonnoli}, G., {Foschini}, L., \& {Maraschi}, L. 2011, \aap, 534, A86

\bibitem[{{Tavecchio} \& {Ghisellini}(2015)}]{tg15}
{Tavecchio}, F. \& {Ghisellini}, G. 2015, \mnras, 451, 1502

\bibitem[{{Tluczykont} {et~al.}(2010){Tluczykont}, {Bernardini}, {Satalecka},
  {Clavero}, {Shayduk}, \& {Kalekin}}]{ltcurves}
{Tluczykont}, M., {Bernardini}, E., {Satalecka}, K., {Clavero}, R., {Shayduk},
  M., \& {Kalekin}, O. 2010, \aap, 524, A48

\bibitem[{van~der Walt {et~al.}(2011)van~der Walt, Colbert, \&
  Varoquaux}]{numpy}
van~der Walt, S., Colbert, S.~C., \& Varoquaux, G. 2011, Computing in Science
  Engineering, 13, 22


\bibitem[{{Wakely} \& {Horan}(2008)}]{tevcat}
{Wakely}, S.~P. \& {Horan}, D. 2008, International Cosmic Ray Conference, 3,
  1341

\bibitem[{{Waxman}(2015)}]{waxman15}
{Waxman}, E. 2015, ArXiv.org, 1511.00815

\bibitem[{{Waxman} \& {Bahcall}(1997)}]{wb97}
{Waxman}, E. \& {Bahcall}, J. 1997, Physical Review Letters, 78, 2292

\bibitem[{{Weekes} {et~al.}(2002){Weekes}, {Badran}, {Biller}, {Bond},
  {Bradbury}, {Buckley}, {Carter-Lewis}, {Catanese}, {Criswell}, {Cui},
  {Dowkontt}, {Duke}, {Fegan}, {Finley}, {Fortson}, {Gaidos}, {Gillanders},
  {Grindlay}, {Hall}, {Harris}, {Hillas}, {Kaaret}, {Kertzman}, {Kieda},
  {Krennrich}, {Lang}, {LeBohec}, {Lessard}, {Lloyd-Evans}, {Knapp},
  {McKernan}, {McEnery}, {Moriarty}, {Muller}, {Ogden}, {Ong}, {Petry},
  {Quinn}, {Reay}, {Reynolds}, {Rose}, {Salamon}, {Sembroski}, {Sidwell},
  {Slane}, {Stanton}, {Swordy}, {Vassiliev}, \& {Wakely}}]{wbbveritas2002}
{Weekes}, T.~C. {et al.}\  2002, Astroparticle Physics, 17, 221

\end{thebibliography}


\end{document}